\documentclass{article}
\usepackage{amsmath}
\usepackage{caption}
\usepackage{indentfirst}
\usepackage{booktabs}
\usepackage{multirow}
\usepackage{float}
\newtheorem{rmk}{Remark}

\usepackage{titlesec}
\titlespacing*{\section}{0pt}{\dimexpr\baselineskip-20pt}{\dimexpr\baselineskip-20pt}

\usepackage{PRIMEarxiv}
\usepackage[utf8]{inputenc} 
\usepackage[T1]{fontenc}    
\usepackage{hyperref}       
\usepackage{url}            
\usepackage{booktabs}       
\usepackage{amsfonts}       
\usepackage{nicefrac}       
\usepackage{microtype}      
\usepackage{lipsum}
\usepackage{fancyhdr}       
\usepackage{graphicx}       
\graphicspath{{media/}}     

\title{Require Process Control? LSTMc is all you need!

}

\author{
  Niranjan Sitapure \\
  Dept. of Chemical Engineering \\
  Texas A\&M University \\
  College Station, TX 77801\\
  \texttt{niranjan\_sitapure@tamu.edu} \\
   \And
  Joseph Sang-Il Kwon$^{*}$\\
  Dept. of Chemical Engineering \\
  Texas A\&M University \\
  College Station, TX 77801\\
  \texttt{kwonx075@tamu.edu} \\
}
\begin{document}

\captionsetup[figure]{font=small,skip=-15pt}
\captionsetup[table]{font=small,skip=5pt}

\maketitle

\begin{abstract}
	
Over the past three decades, numerous controllers have been developed to regulate complex chemical processes, including PI/PID controllers and MPC. However, these control approaches have certain limitations. Traditional PI/PID controllers often require customized tuning for various set-point tracking scenarios, as they lack grade-to-grade (G2G) transferability. On the other hand, MPC frameworks involve resource-intensive steps and the utilization of black-box machine learning (ML) models can lead to issues such as local minima and infeasibility. To address these challenges, there is a need for an alternative controller paradigm that combines the simplicity of a PI controller with the G2G transferability of an MPC approach. In this study, we introduce the novel concept of an LSTM controller (LSTMc) as a model-free data-driven controller framework. The LSTMc considers an augmented input tensor that incorporates information on state evolution and error dynamics for the current and previous $W$ time steps, to predict the manipulated input at the next step ($u_{t+1}$). To demonstrate the proposed framework, batch crystallization of dextrose was taken as a representative case study. The desired output for set-point tracking was the mean crystal size ($\bar{L}$), with the manipulated input being the jacket temperature ($T_j$). Extensive training data, encompassing 7000+ different operating conditions, was compiled, including various cooling curves for $T_j$, seeding conditions, and varying initial concentrations, to ensure comprehensive training of LSTMc across a wide state space region. For comparison, we also designed a PI controller and an LSTM-MPC for different set-point tracking cases. The results consistently showed that LSTMc achieved the lowest set-point deviation ($<$2\%), which was three times lower than that of the MPC. Remarkably, LSTMc maintained this superior performance across all set-points, even when sensor measurements contained noise levels of 10\% to 15\%. In summary, by effectively leveraging process data and utilizing sequential ML models, LSTMc offers an alternative controller design approach. The results demonstrate its superiority as a controller in terms of set-point tracking accuracy and transferability, making it a promising solution for complex chemical processes.
\end{abstract}

\section{Introduction}

\setlength{\belowcaptionskip}{10pt}
\setlength{\parindent}{15pt} 

Control of complex chemical processes (e.g., crystallization, fermentation, battery dynamics, hydraulic fracking, etc.) is a ubiquitous and high-value endeavor in the industry, which often requires specially tuned controllers or advanced model-based control techniques \cite{Su2009Adaptive, siddhamshetty2018model, braatz2002advanced, nagy2012advances, torchio2015real, hwang2022model, sitapure2022neural}. Generally, the literature suggests three main control approaches for such systems. The first approach involves the use of proportional-integral (PI) or PI-differential (PID) controllers, which are specifically tuned for set-point tracking under well-defined process conditions. Despite the simplicity and reliable performance of PI/PID controllers, they require custom tuning for different set-points and widely varied operating conditions \cite{rohani1990self}. Second, to address these issues, model predictive controllers (MPC) have been demonstrated that utilize a combination of the state-space model of the chemical system, and an internal optimization formulation to predict the future state evolution of the state and take corrective action in the form of manipulated inputs. Although MPC has certain advantages (e.g., the inclusion of explicit process constraints, multi-objective control, simultaneous set-point tracking, disturbance rejection, and incorporating state estimation), it requires a computationally inexpensive state-space model that accurately describes the system dynamics. Unfortunately, for the case of complex chemical systems, a simple state-space model is not enough, and often high-fidelity models are utilized \cite{SITAPURE2020127905, sitapure2021cfd, kwon2014modeling,kwon_PECVD}. Third, as these models cannot be directly incorporated within the MPC, various data-driven surrogate modeling techniques are utilized. For example, Sandoval and colleagues demonstrate using a deep neural network (DNN)-based surrogate model for implementation of MPC in thin-film deposition process \cite{kimaev2019nonlinear}. Similarly, Kwon and colleagues have demonstrated the use of a long-short-term-memory (LSTM) network within an MPC framework for accurate control of a batch fermentation process \cite{shah2023achieving}. Along the same lines, Wu and colleagues have demonstrated various forms of recurrent neural networks (RNNs) to mimic the complex dynamics of a batch crystallizer, and then incorporate them within an MPC to perform a set-point tracking task \cite{zheng2022machine,zheng2022online}. Further, sparse identification of system dynamics (SINDy) and operable adaptive sparse identification of systems (OASIS)-based models have been utilized to develop surrogate models that can be integrated with an MPC framework for regulating complex chemical processes \cite{bhadriraju2019machine, bhadriraju2020operable}. 

Despite their reliable performance, the abovementioned control approaches have certain limitations: (a) Traditional PI controllers show poor grade-to-grade (G2G) transferability \cite{vilanova2010analysis}, often requiring bespoke tuning for different set-point tracking cases \cite{pannocchia2005candidate}; (b) Utilization of an MPC framework entails multiple resource-intensive steps (i.e., training and testing of a surrogate model, formulation of an internal optimization problem, and tuning of the MPC); and (c) frequently, the use of black-box- based machine-learning (ML) models in MPC can lead to complications such as navigating through regions of infeasibility, nonconvexity, and local minima. Moreover, existing industrially available controller hardware does not have the computational bandwidth for quick online computations required by an MPC, thereby hindering their practical use in chemical operations. Therefore, it is critical to recognize the need for an alternative approach in controller implementation. Ideally, this innovative controller framework should have the ability to consider both state dynamics, like an MPC, and error dynamics, similar to a PI controller, concurrently. Such a holistic approach could potentially address the aforementioned challenges more effectively.   

To this end, we can draw inspiration from the LSTM network, which has certain unique characteristics that result in the consistently superior performance of LSTM for time-series tasks \cite{wang2023attention}. Specifically, these advantages can be attributed to three key characteristics \cite{\cite{hochreiter1997long}}. First, LSTM networks are sequential ML models that explicitly consider multidimensional time-series data as compared to DNNs or CNNs. Second, the LSTM network is equipped with four distinct internal mechanisms, viz., a memory cell, forget gate, input gate, and output gate. These mechanisms compute the contextual relevance between current and previous state information. The sequential activation of these gates allows only pertinent state information to be passed onto the next time-step. As a result, LSTMs learn to highlight significant process changes by assigning more weight to these time steps and ignoring weak process changes. This characteristic is particularly suited for process control tasks, as LSTM networks can prioritize the evolution of the process state when a significant control action is implemented at a certain time step $t$. Third, since each of these successive gates computes the contextual relevance between current and previous time-steps using specially trained \textit{sigmoid} functions, a noisy process signal gets dampened and does not hamper the model predictions. Given these attributes, LSTM networks emerge as strong candidates for developing a model-free, data-driven LSTM-controller (i.e., LSTMc) framework. This innovative approach utilizes state feedback and error information from not just the current time-step, but also from previous $W$ time steps to learn an offline complex control law. This law is applicable across the entire state-space and demonstrates high G2G transferability. Essentially, these attributes enable the LSTMc to (a) understand the relationship between state evolution and error dynamics, (b) use the weighting mechanism of internal gates to emphasize significant process changes (e.g., control actions), and (c) filter out noisy process measurements to provide accurate predictions for manipulated inputs at the next time-step. These abilities allow the LSTMc to drive the system towards the desired set-point during closed-loop operation in a superior manner compared to traditional controllers. 

To demonstrate the proposed framework, batch crystallization of dextrose was used as a representative case study, typifying a complex, nontrivial chemical system. Within this context, the jacket temperature ($T_j$) serves as the manipulated input, while the mean crystal size ($\bar{L}$) acts as the desired output requiring set-point tracking. Consequently, a comprehensive compilation of open-loop training data ($\mathbb{D}$) was generated by simulating the process for 7000+ different operating conditions (i.e., different cooling curves for $T_j$, seeding conditions, and varying initial concentration) to ensure encompassing a large state-space. First, an LSTMc was trained using an augmented input tensor (i.e., state information and error values) for current and previous $W$ time-steps to learn a unified control law that correlates the augmented input tensor with the jacket temperature at the next time-step ($T_j(t+1)$) to yield the desired set-point ($\bar{L}_{sp}$). Second, an MPC that utilizes a surrogate model of the batch crystallization process was formulated for the same task of set-point tracking. Interestingly, the surrogate model within the LSTM-MPC is another LSTM model, which has the same architecture and number of parameters ($N_p$), and is trained using the same dataset ($\mathbb{D}$) to predict state evolution of the crystallization process. Third, a set of PI controllers, each with bespoke tuning parameters for specific set-points, were also developed. Comparison of the controller performance for the three controllers highlights that the LSTMc consistently outperforms the others, exhibiting a set-point deviation of less than 2\% across all cases. This precision is threefold better than the LSTM-MPC's performance. Furthermore, in terms of internal computations, the LSTMc proves to be remarkably efficient, operating 2000 times faster. Moreover, while a PI controller accurately tuned for a particular set-point may show a set point deviation of less than 2\%, its performance diminishes in comparison to the LSTMc when tested in another case, demonstrating a set-point deviation twice as large. Impressively, the LSTMc consistently maintains a high standard of set-point tracking performance across all varying set-points, even when the noise in sensor measurements ranges from 10 to 15\%. In a nutshell, LSTMc introduces an alternative approach to controller design. It adeptly leverages the availability of process data and the efficient use of sequential ML models, yielding a controller with superior performance. 

The rest of this manuscript is organized as follows: we begin with a concise mathematical representation of batch crystallization, which acts as a representative case study. This is followed by an exploration of the internal computations of the LSTMc, LSTM-MPC, and PI controller. Then, model validation and controller results are presented. The paper concludes with a discussion and summary of the findings from our research.

\section{Mathematical Modeling}
\subsection{Batch Crystallization of Dextrose}

As mentioned earlier, batch crystallization of dextrose is considered as a representative case study to showcase the capabilities of the LSTMc when applied to a complex, nontrivial chemical system. To this end, the crystal growth rate and nucleation rate serve as primary descriptors of the kinetics of a crystal system. For dextrose, the growth rate ($G$) is given as follows \cite{markande2012influence}:

\begin{equation}
	\begin{split}
		G~(m/s) = 1.14 \times 10^{-3}exp\left(\frac{-29549}{RT}\right) \sigma^{1.05}
	\end{split}
\end{equation}
where $R$ is the universal gas constant, $T$ is the temperature, and $\sigma$ is the relative supersaturation ratio.  In the specific context of this dextrose crystallization study, seed crystals are introduced at $t=0$ with an average crystal size ($\bar{L}$) ranging from 100 to 125 $\mu$m and a standard deviation of 5 to 25$\mu$m. Due to the large mass of seed crystals (i.e., suspension density, $M_T$) and the shearing of growing crystals induced by the agitator, significant heterogeneous nucleation is noted. Thus, the nucleation rate ($B$) is established as follows \cite{markande2012influence}:

\begin{equation}
	\begin{split}
		B~(\#/kg \cdot s) = 4.50 \times 10^{4}M_T^{0.49}(\sigma)^{1.41}
	\end{split}
\end{equation}
Next, the size distribution of crystals and their temporal evolution can be traced using the population balance model (PBM), which utilizes a population density function, $n(L,t)$, and is given as follows \cite{worlitschek2004model}:

\begin{equation}
	\begin{aligned}
		\frac{\partial n (L,t)}{\partial t} + \frac{\partial (G(T,C_s)n(L,t))}{\partial L} = B(T,C_s)
	\end{aligned}
\end{equation}
where $n (L,t)$ represents the number of crystals of size $L$ at time $t$, $B(T,C_s)$ is the total nucleation rate, and $G(T,C_s)$ represents the crystal growth rate. Then, the PBM is integrated with mass and energy balance equations, which are presented below:
\begin{equation}\label{eq1}
	\begin{gathered}
		\frac{d C_s}{d t} = -3\rho_ck_vG\mu_2 \\
		mC_p\frac{dT}{dt} = -UA(T - T_{j}) - \Delta H\rho_{c}3k_vG\mu_2
	\end{gathered}
\end{equation}
where $\mu_2$ is the second moment of crystallization, $k_v$ is the shape factor, $\rho_c$ is the crystal density, $C_p$ is the heat capacity of the crystallization slurry, $m$ is the total mass of the slurry, $U~A$ is the area-weighted heat transfer coefficient, and $\Delta H$ is the heat of crystallization. For a more comprehensive description of modeling crystallization systems, the reader is referred to relevant literature sources \cite{kwon2013modeling_batch,kwon2013modeling_aggregate,kwon2014crystal,sitapure2023unified, ochsenbein2014crystallization}.

\subsubsection{Data Generation} 

\begin{figure}[!ht]
	\begin{center}
		\centerline{\includegraphics[width=1\columnwidth]{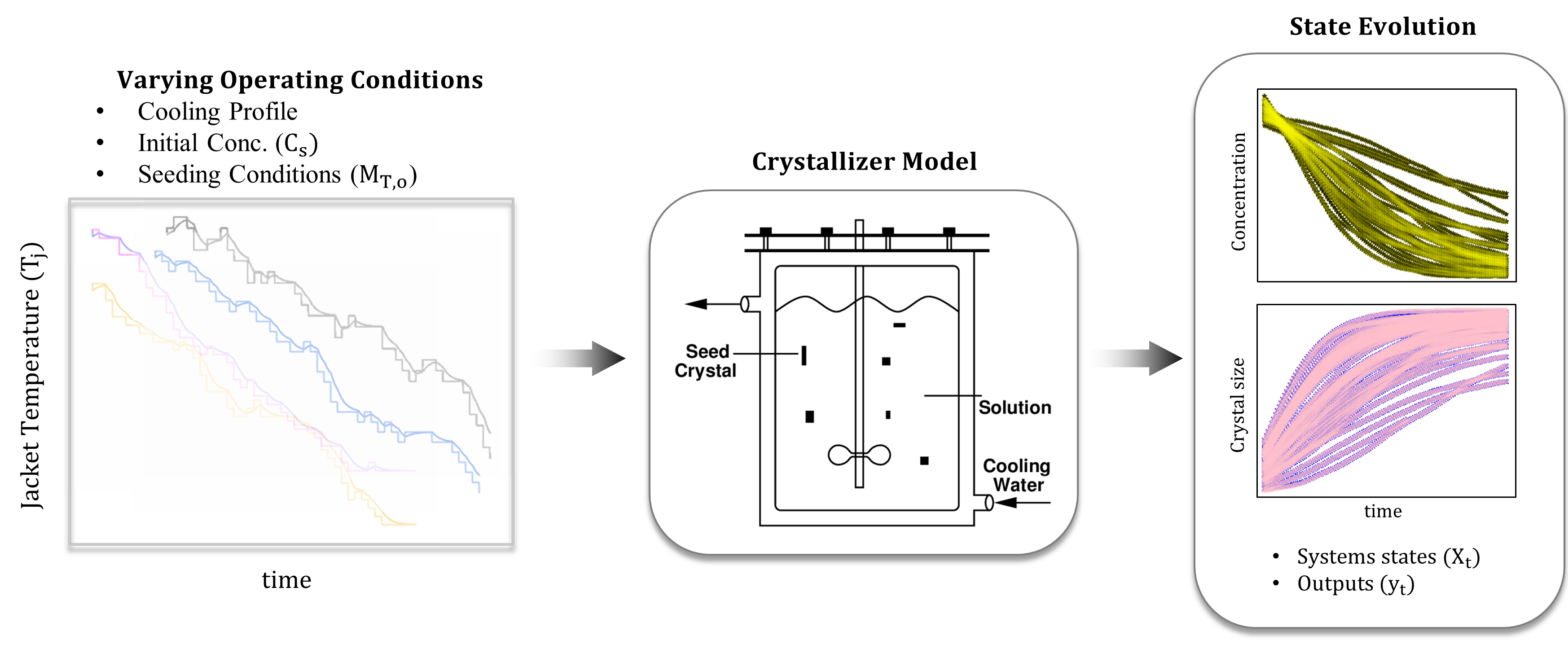}}
	\end{center}
	\caption{A schematic illustration of simulating the batch crystallization of dextrose. Here, data for only 1000 operating conditions is shown to avoid overcrowding of the results. }
	\label{crystallization_schematic}
\end{figure}

The PBM and mass and energy balance equations can be solved using native Python solvers for a specified temperature curve of the cooling jacket, and the temporal evolution of crystal size, crystallizer temperature, concentration, and other system states can be acquired as shown in Figure~\ref{crystallization_schematic}. Consequently, 7000 different step cooling curves were randomly generated to encompass a large set of possible operating conditions for the batch crystallizer. More precisely, for each cooling profile, a sampling time of 60 min. was considered, and the maximum temperature change between two sampling times ($\Delta T$) was limited to $\pm$ 7$^{\circ}$C. Next, the PBM was simulated for all 7000 cases, and a state matrix ($X_t$) was recorded for a total operating time of 24 hours. Also, the cooling profiles were bounded between a certain temperature range (i.e., [5, 45]$^{\circ}$C), while the initial solute concentration was varied between 0.55 to 0.75 kg/kg. As a result, the data was divided into 350K data points for the training set, 150K for the validation set, and another 100K for the testing set. Each data point contains information comprising of 9 process states (i.e., [$T_j, C_s, T, \bar{L}, \mu_0, \mu_1, \mu_2, \mu_3, t$]) for both the current and previous $W$ time-steps.  

\subsection{RNN and LSTM Architecture}
\subsection{RNN Models}
\begin{figure}[!ht]
	\begin{center}
		\centerline{\includegraphics[width=0.65\columnwidth]{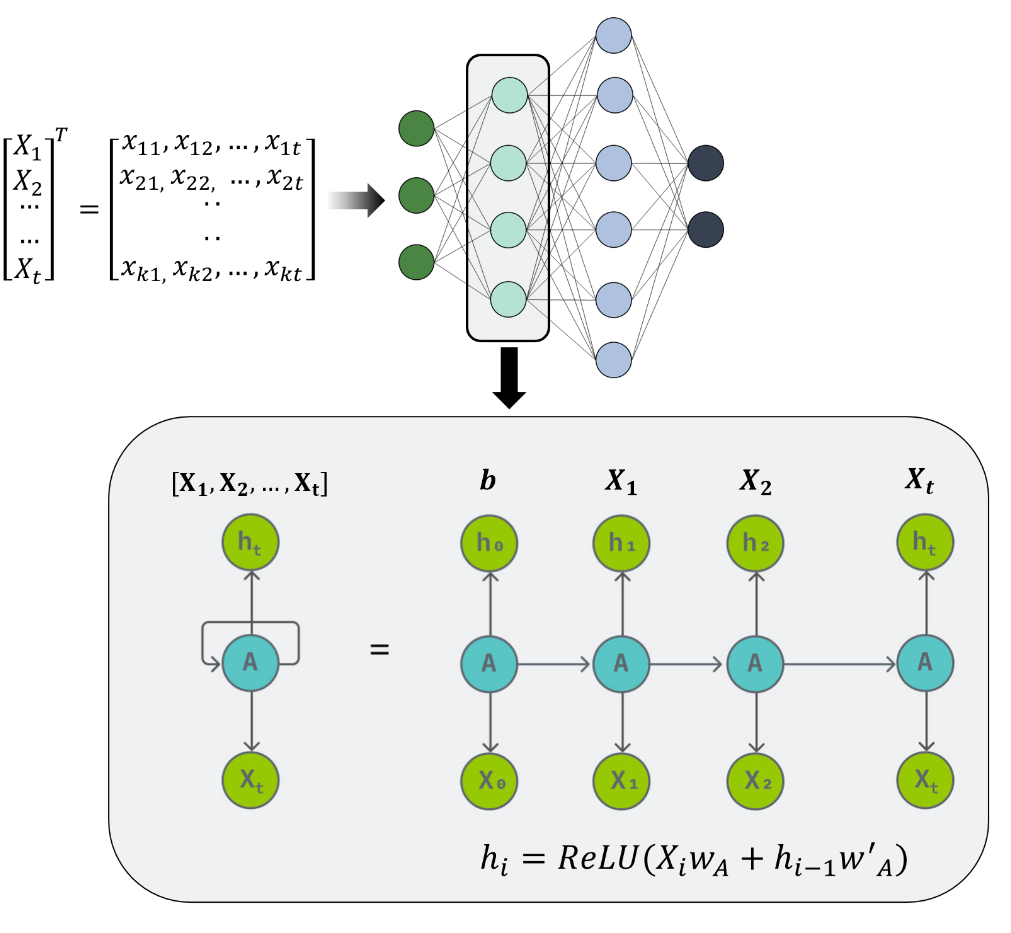}}
	\end{center}
	\caption{A schematic illustration of an RNN.}
	\label{RNN_schematic}
\end{figure}

The LSTM is a special type of RNN model, and thus, for understanding the working of an LSTM, a brief overview of RNNs is required. First, as shown in Figure~\ref{RNN_schematic}, an RNN takes in a $k$-dimensional input tensor consisting of temporal information spanning $t$ time-steps (i.e., $[X_1, X_2, ..., X_t]$). A multilayered RNN then processes these inputs sequentially through layer $A$, which contains $\lambda$ number of neurons, resulting in hidden states (i.e., $[h_1, h_2, ..., h_t]$). It is noteworthy that the computation at any time step $i$ employs an activation function, $\phi$ (in this case, the Rectified Linear Unit (ReLU) that processes the input information ($X_i$) and the hidden state from the previous time-step ($h_{i-1}$), each of which is multiplied by their respective weights $w_A$ and $w^{\prime}_A$. In summary, for an RNN layer $A$, the internal computations can be presented as follows: 

\begin{equation}\label{RNN_equations}
\begin{gathered}
	h_1 = \phi \left(X_1 w_A + h_0 w^{\prime}_A \right) \\
	h_2 = \phi \left(X_2 w_A + h_1 w^{\prime}_A \right) \\
	...\\
	h_i = \phi \left(X_i w_A + h_{i-1} w^{\prime}_A \right)
\end{gathered}
\end{equation}

Other RNN layers adhere to a similar computational structure, albeit with different network weights (e.g.,  $w_B$, $w_C$, and so on) \cite{rumelhart1985learning}. Although RNNs work well when only a few time-steps are included in the input tensor, a large number of time-steps can lead to issues of vanishing or exploding gradients. This arises from the convolution of activation functions across each previous hidden state. For example, if there are 3 total time-steps in the source sequence, $h_3$ can be expressed as: 

\begin{equation}\label{RNN_convolution}
	\begin{gathered}
		h_3 = \phi \left(X_3 w_A + h_2 w^{\prime}_A \right) \\
		h_3 = \phi \left(X_3 w_A + \left(\phi \left(X_2 w_A + h_1 w^{\prime}_A \right)\right) w^{\prime}_A \right) \\
		h_3 = \phi \left(X_3 w_A + \left(\phi \left(X_2 w_A + \left(\phi \left(X_1 w_A + h_0 w^{\prime}_A \right)\right) w^{\prime}_A \right)\right) w^{\prime}_A \right)
	\end{gathered}
\end{equation}
During backpropagation in an RNN, the convolution functions described in Equation~\ref{RNN_convolution} can result in vanishing gradients if the activation function is a \textit{sigmoid}, and exploding gradients if the activation function is an \textit{ReLU}. This issue becomes more pronounced as the number of time-steps in the input tensor increases \cite{hu2018overcoming}. Unfortunately, in complex chemical processes where an adequate number of time-steps (10 or more) are required in the input tensor to provide substantial information about the state evolution, the issue of vanishing or exploding gradients impedes training performance. 

\begin{figure}[!ht]
	\begin{center}
		\centerline{\includegraphics[width=1\columnwidth]{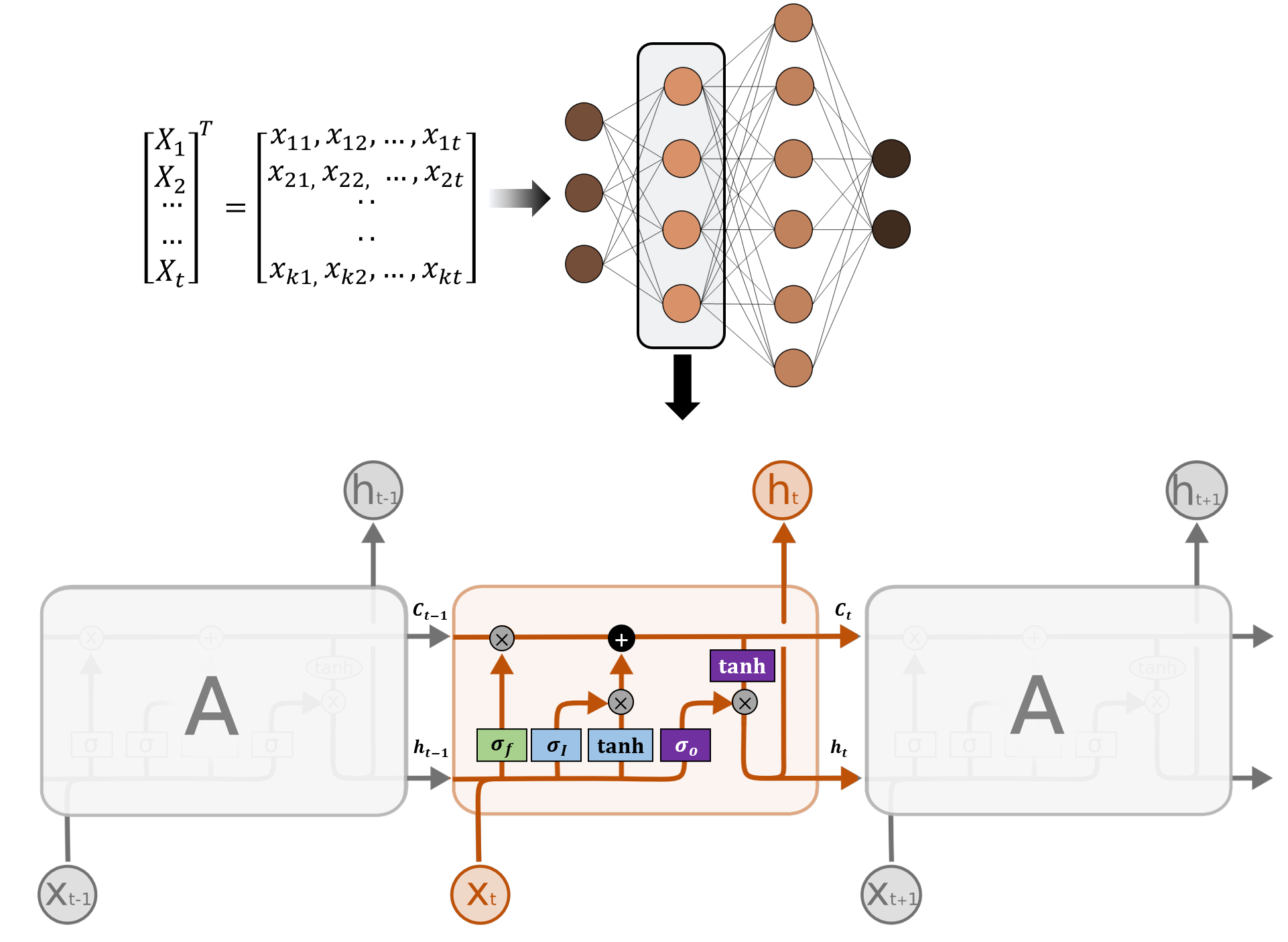}}
	\end{center}
	\caption{A schematic illustration of an LSTM.}
	\label{LSTM_schematic}
\end{figure}

\subsection{LSTM: A Special Type of RNN}

To tackle the above issue, LSTM networks were introduced \cite{hochreiter1997long}. LSTMs efficiently manage memory to overcome the issue of vanishing or exploding gradients, and they also provide accelerated training times. An LSTM network (Figure~\ref{LSTM_schematic}) has similarities with an RNN (Figure~\ref{RNN_schematic}) in terms of layers, hidden states, and sequential computation of time-steps. However, it also includes four additional internal components: a memory cell ($C_t$), forget gate ($\sigma_f$), input gate ($\sigma_I$), and output gate ($\sigma_O$). At every time-step $t$, the LSTM uses the memory cell and hidden state from the previous time-step $t-1$ to conduct a series of internal computations as shown below \cite{wang2023attention}:

\begin{equation}\label{LSTM_equations}
	\begin{gathered}
		f_t = \sigma_f\left(w_f\cdot[h_{t-1}, X_t] + b_f\right) \\
		I_t = \sigma_I\left(w_I\cdot[h_{t-1}, X_t] + b_I\right) \\
            \bar{C_t} = tanh\left(w_C\cdot[h_{t-1}, X_t] + b_C\right) \\
		C_t = f_t\cdot C_{t-1} + I_t\cdot \bar{C_t} \\
            O_t = \sigma_O\left(w_O\cdot[h_{t-1}, X_t] + b_O\right) \\
		h_t = O_t\cdot tanh(C_t) 
	\end{gathered}
\end{equation}
where $(\sigma_f, w_f, b_f)$, $(\sigma_I, w_I, b_I)$, $(\sigma_O, w_O, b_O)$ are the sigmoid function-based neural networks, weights, and bias of the forget gate, input gate, and output gate, respectively. During forward propagation in an LSTM, the augmented tensor $[X_t, h_{t-1}]$ is first processed by the forget gate, which uses $\sigma_f$ to compute the contextual difference between the current and previous time-steps. Next, the augmented tensor $[h_{t-1}, X_t]$ is fed through the input gate, which determines the relevant information ($I_t$) to be processed by the LSTM. As the previous steps have erased some of the irrelevant information from the time-steps, new pertinent information must be added to the memory cell ($C_t$). This is accomplished by combining the modified memory cell from the previous time-step ($C_{t-1}$) with the current internal memory cell ($\bar{C_t}$). Finally, the output gate ($\sigma_O$) controls the amount of relevant information to be passed on to the next LSTM cell and calculates the current hidden state ($h_t$). 

The above computations (Equation~\ref{LSTM_equations}) are conducted iteratively for each preceding time-step, which preserves the temporal information of the source sequence intact. More importantly, the combination of the forget, input, and output gates allows the LSTM to adeptly attend to substantial, pertinent changes over the long-term, such as control actions, while filtering out irrelevant short-term changes like process noise. Essentially, these various gates enable the LSTM to assign different weights to each of the time-steps in the source sequence. For example, a time-step wherein a control action was implemented may be given a higher weight, while a time-step with no significant change in state evolution might receive a lower weight. This method contrasts sharply with RNNs, which give equal weights to all time-steps. The internal gates of the LSTM consequently also act as a dampening system, or an internal noise filter, that learns to focus on significant process signal changes. Specifically, if an entire input tensor consists of 50 time-steps, including 5 control actions and some process noise, an LSTM can selectively assign more weight to system dynamics around these 5 control actions while concurrently filtering out the process noise. Moreover, during LSTM training, these internal gates, along with their respective weights and bias functions, are optimized to ensure that the network learns an ideal weighting scheme, thereby fostering generalization across the entire state-space. Given these features, LSTMs are an ideal candidate for mimicking a process controller, as they can (a) adapt to changing input or feedback signals, and (b) efficiently filter out process noise, which is ubiquitous in industrial settings.

\begin{figure}[!ht]
	\begin{center}
		\centerline{\includegraphics[width=1\columnwidth]{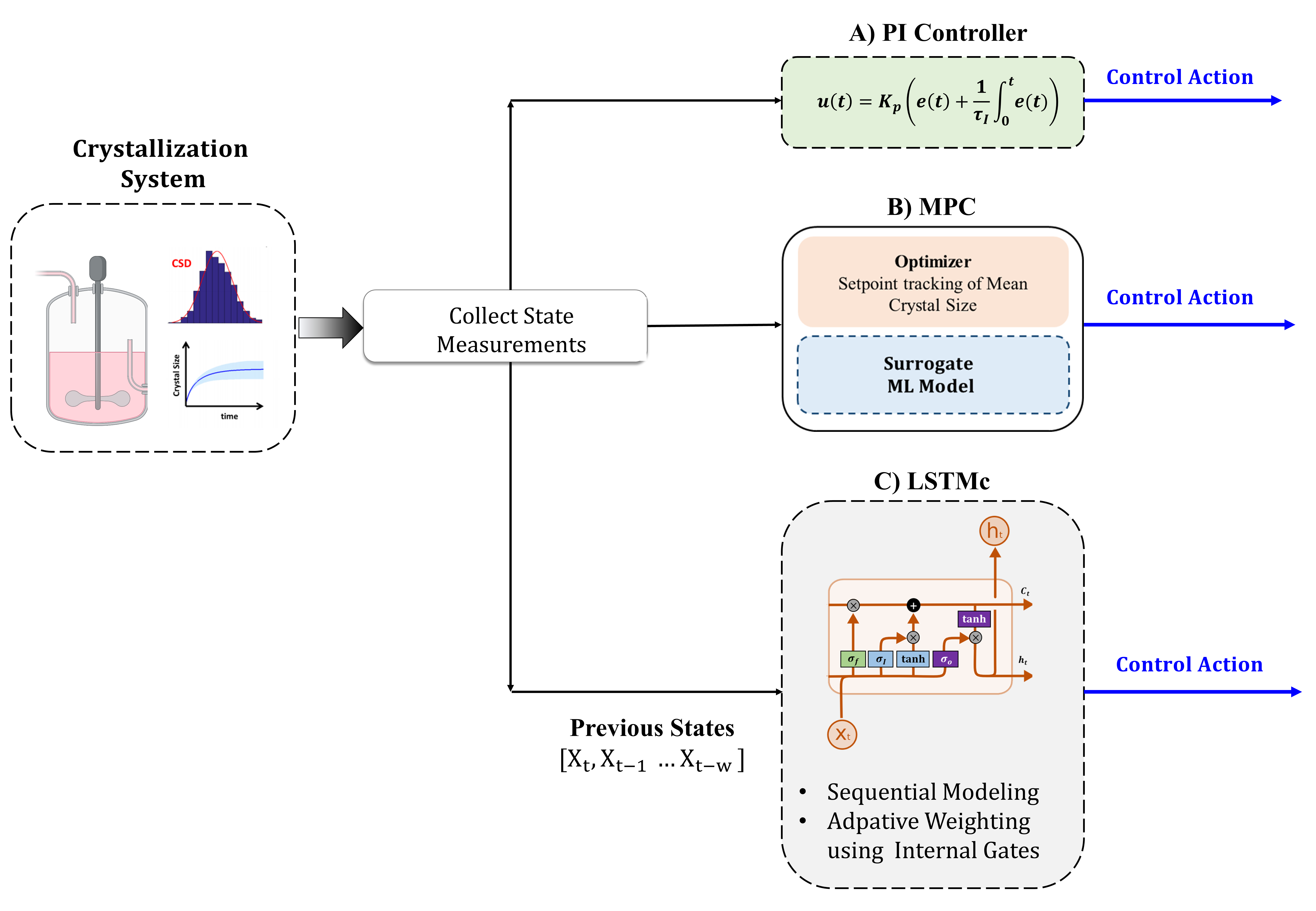}}
	\end{center}
	\caption{A schematic illustration of different controllers considered in this work including (a) PI controller, (b) LSTM-MPC, and (c) LSTMc.}
	\label{schematic_controller_comparison}
\end{figure}

\section{Controller Design}

As mentioned earlier, this work aims to develop and demonstrate a model-free, data-driven controller named long-short-term-memory controller (LSTMc). The LSTMc strives to mimic the operation of LSTM networks by predicting future input profiles to steer the system toward a desired value. More precisely, it takes into account the state evolution and error dynamics of the current and previous $W$ steps to predict the input value at the next time-step (i.e., $T_j(t_{k+1})$) that will drive the system towards the set-point in a closed-loop operation. In order to evaluate the LSTMc's performance in comparison with state-of-the-art approaches, we employed two well-known approaches: the MPC and PI controller. The LSTMc was developed using the simulation data produced in the preceding section. Next, we implemented an LSTM-MPC framework for a batch crystallizer, leveraging a separate LSTM model as an ML-based surrogate model. It is important to underscore that the LSTM model employed within the LSTM-MPC framework is different from the LSTMc. While the former provides a time-series prediction of system states, the latter is designed to predict the next input action. Lastly, we developed a PI controller and fine-tuned it for specific set-point tracking scenarios in the batch crystallizer. A schematic comparison of these three controllers is shown in Figure~\ref{schematic_controller_comparison}.

\begin{figure}[!ht]
	\begin{center}
		\centerline{\includegraphics[width=0.85\columnwidth]{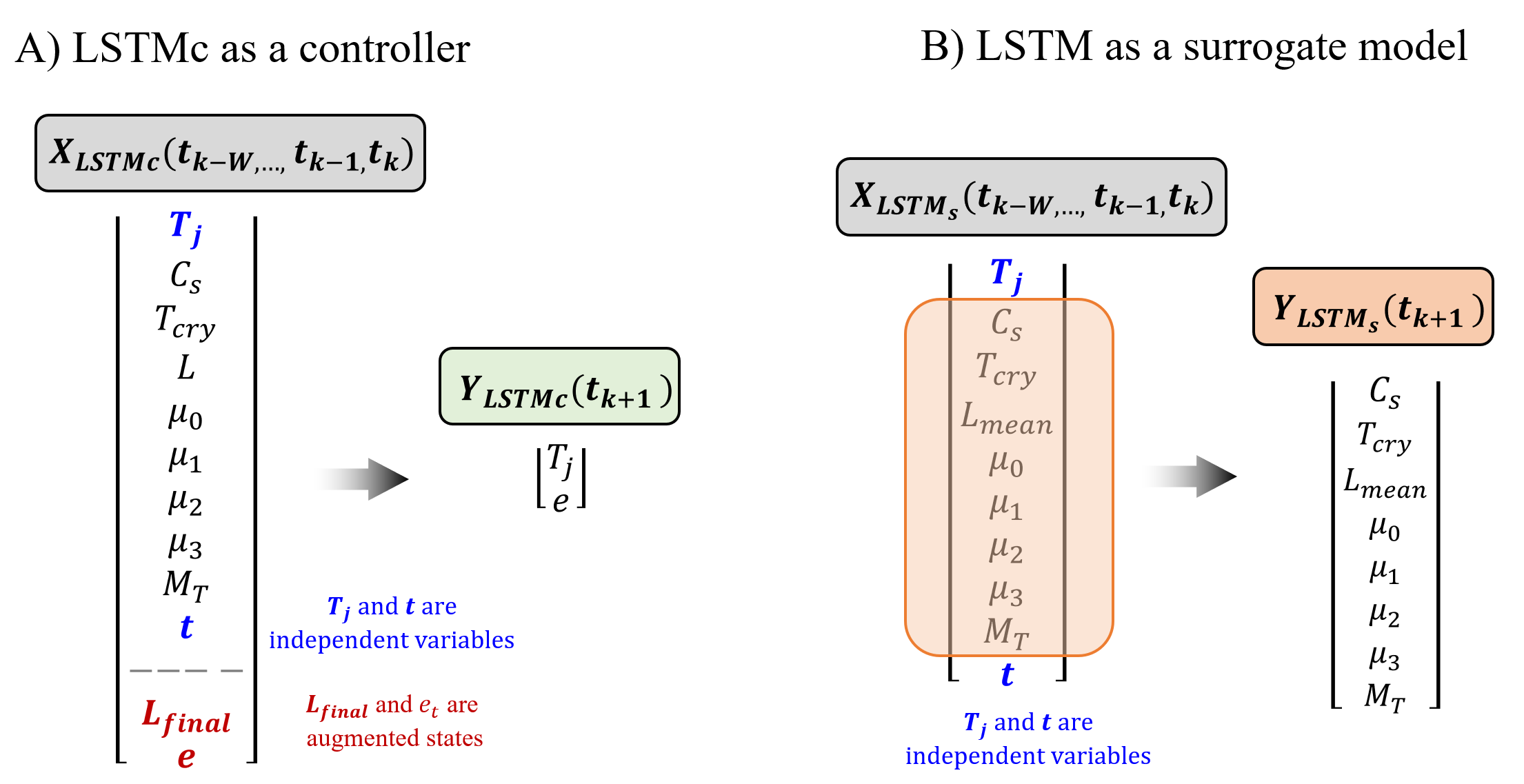}}
	\end{center}
	\caption{A schematic illustration of the input and output of (a) LSTMc, and (b) LSTM-based surrogate model.}
	\label{LSTMc_vs_LSTM}
\end{figure}

\subsection{Designing LSTMc}
\begin{figure}[!ht]
	\begin{center}
		\centerline{\includegraphics[width=0.5\columnwidth]{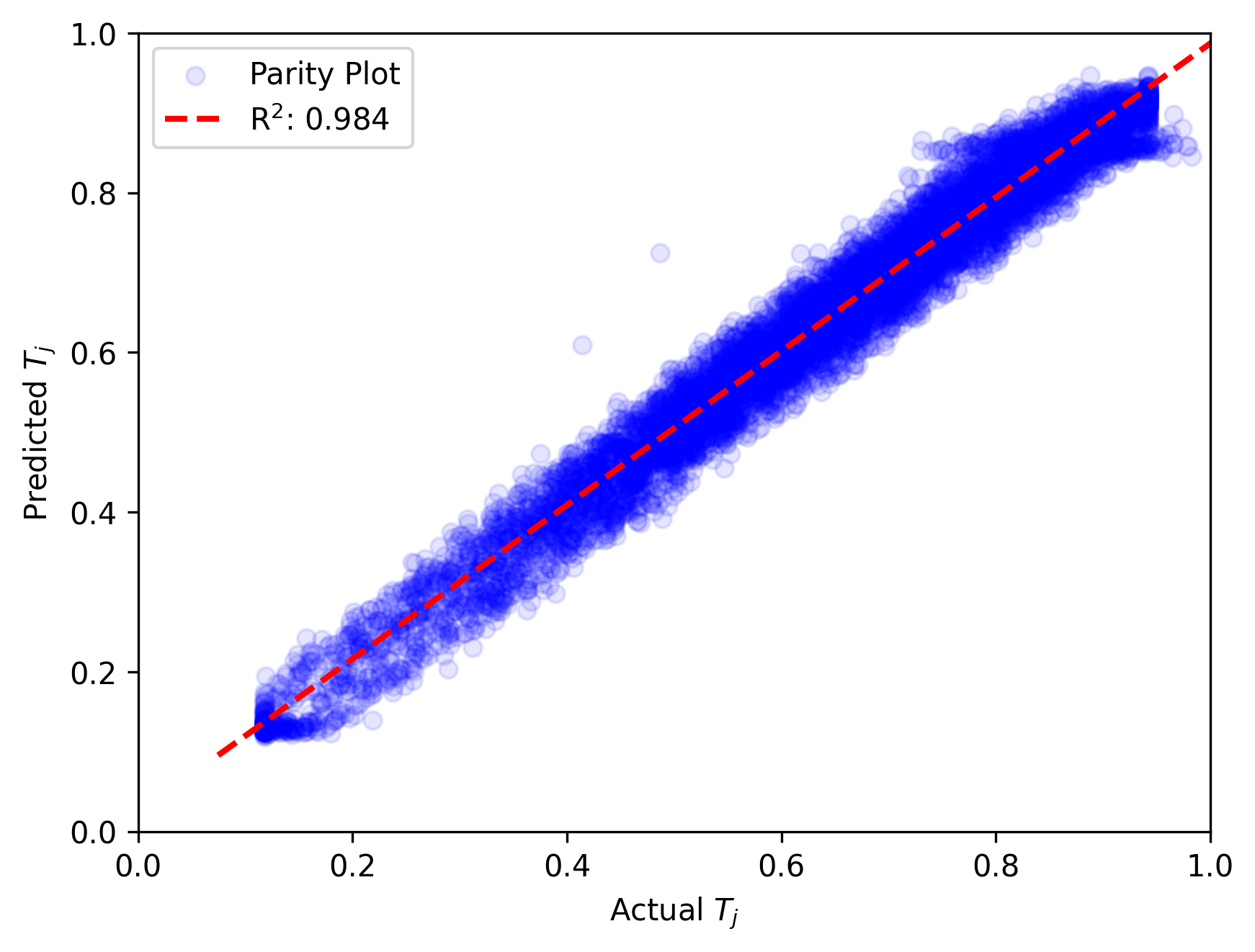}}
	\end{center}
	\caption{The parity plot illustrating the level of agreement between the LSTMc predictions and the simulated $T_j$ values for the testing dataset.}
	\label{LSTMc_parity_plot}
\end{figure}

In the case of LSTMc, an augmented tensor (i.e., [$T_j, C_s, T, \bar{L},\mu_0,\mu_1,\mu_2,\mu_3, M_T, t~|~{L_{final}}, e$]) was utilized as the input to generate an output tensor $Y_{LSTMc} = [T_j(t_{k+1}), e(t_{k+1})]$, as shown in Figure~\ref{LSTMc_vs_LSTM}a. The term $L_{final}$ represents the terminal crystal size obtained by simulating the crystallization process under particular operating conditions, such as an arbitrarily selected cooling profile $T_j$, a particular initial solute concentration ($C_{s,o}$), and seeding condition ($M_{T,o}$)). The error $e$ is defined as $e = \left(L_{final}-\bar{L}\right)^2$. Three key aspects underline the rationale for using such an augmented input tensor. First, the training data, produced in an open-loop manner, comprises more than 7000 unique operating conditions, each with its own jacket temperature profile, initial solute concentration, and seeding conditions. Consequently, each of these conditions triggers a unique state evolution of the crystallizer (as shown in Figure~\ref{crystallization_schematic}), leading to a distinct terminal crystal size at $t=24$ hours (i.e., ${L_{final}}$). Incorporating ${L_{final}}$ as an augmented state provides the LSTMc with a contextual understanding of the terminal state of the crystallization process, which is fundamentally the goal of a set-point tracking task. Second, the addition of state errors $e$ for the current and previous $W$ time-steps enables the LSTMc to learn the error dynamics. To minimize the error at the next time-step, the LSTMc needs to understand and predict error dynamics under diverse operating conditions. Hence, the inclusion of $e$ as an augmented state is necessary. Third, the state information for the current and previous $W$ time-steps (i.e., $[T_j, C_s, T, \bar{L}, \mu_0, \mu_1, \mu_2, \mu_3, M_T, t]$) enables the LSTMc to learn the correlation between system states and the resulting error. In simpler terms, the integration of system states and error dynamics provides the LSTMc with context regarding how the current state influences the error at the next time-step, and what value of manipulated input will minimize this error. Employing this approach, the LSTMc was developed using a training set comprising 350K data points, a validation set with 150K data points, and a testing set with additional 100K data points. Figure~\ref{LSTMc_parity_plot} shows the comparison between the $T_j$ predictions from the LSTMc and those from the simulated testing dataset. Additionally, the normalized-mean-squared error (NMSE) for the testing dataset stands at $1.08 \times 10^{-3}$, indicating a highly accurate model fit.

The implementation of the LSTMc is similar to the typical closed-loop controller setup. More specifically, at each sampling interval, the LSTMc takes in an augmented state tensor (i.e., [$T_j, C_s, T, \bar{L}, \mu_0, \mu_1, \mu_2, \mu_3, M_T, t, {L_{final}}, e$]) for the current and previous $W$ time-steps. Here, it is important to note that the input to the LSTMc model comprises not only system state information (i.e., $T_j, C_s, T, \bar{L}, \mu_0, \mu_1, \mu_2, \mu_3, M_T, t$) but also includes two additional states (i.e., ${L_{final}}, e$). These additional inputs provide necessary context by detailing (a) the target value at the end of the crystallization (${L_{final}}$), and (b) the deviation of the current state from this target value, as indicated by $e = \left(L_{final}-\bar{L}\right)^2$. This context equips the LSTMc with the information needed to determine the control action for the next time-step, thereby minimizing the deviation error. Consequently, using this augmented state tensor, LSTMc calculates the $T_j$ value that would minimize the discrepancy between ${L_{final}}$ and $\bar{L}$. The predicted $T_j$ is then implemented in the crystallization system, which is the PBM-based batch crystallizer model, serving as a virtual experiment. The evolution of state variables is observed and then fed back into the LSTMc. This cycle is repeated until the end of the crystallization process (i.e., 24 hours). The terminal set-point deviation is then computed, providing a performance metric for LSTMc's set-point tracking ability.  

\subsection{Design of LSTM-MPC}

\begin{figure}[!ht]
	\begin{center}
		\centerline{\includegraphics[width=1\columnwidth]{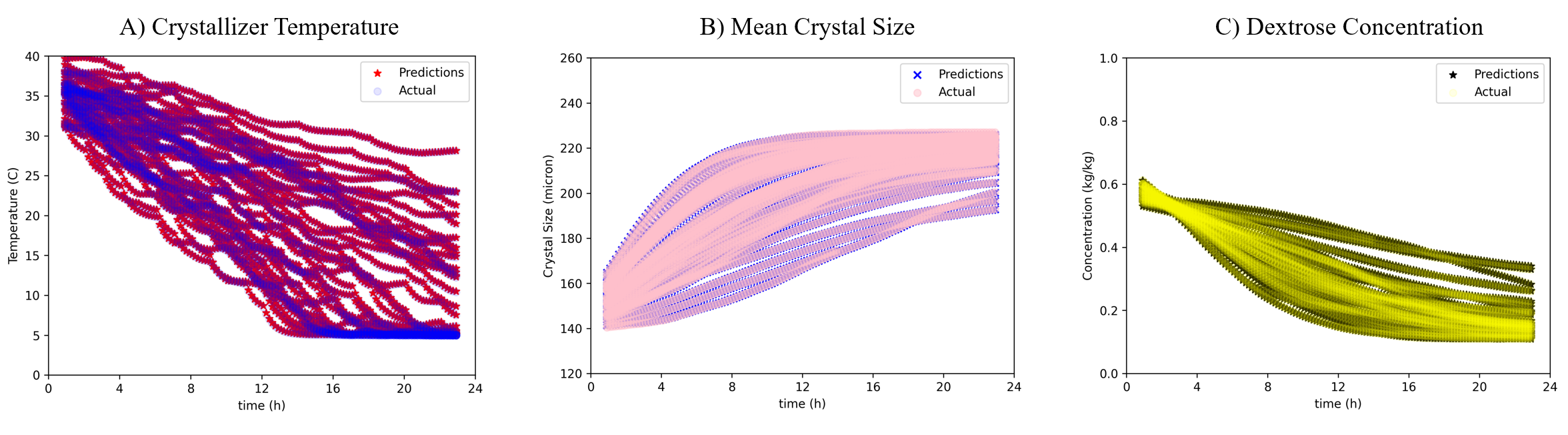}}
	\end{center}
	\caption{The validation results obtained from $LSTM_s$ are presented for the (a) crystallizer temperature, (b) dextrose concentration, and (c) mean crystal size.}
	\label{LSTM_validation}
\end{figure}

Next, we formulated an MPC for the batch crystallization of dextrose, aiming to track the set-point of the mean crystal size ($\bar{L}_{sp}$) by manipulating the jacket temperature ($T_j$). More precisely, at every sampling time, the MPC determines an optimal series of inputs by employing another LSTM network as a surrogate model. However, only the first input from this sequence is implemented and carried forward to the next sampling time. Subsequently, this optimal input is supplied to a virtual experiment, in this case, the PBM for the batch crystallization of dextrose). At the next sampling time, available measurements are fed back into the MPC. Moreover, our proposed MPC operates in a receding horizon fashion. That is, at the initial sampling time, a full trajectory of optimal inputs is calculated (e.g., a 10-step input sequence). As the process moves closer to the end time, fewer optimal inputs are calculated until finally, the last optimal input is computed. An MPC problem was thus formulated following the approach outlined in Figure~\ref{schematic_controller_comparison}b. The goal was to minimize the deviation from a set-point mean crystal size while respecting practical operating constraints, as detailed below:

\begin{equation} \label{optimizer}
	\begin{aligned}
		& \underset{T_{j}(t)}{Minimize}
		&& \left (L- L_{sp} \right)^2  \\
		&\text{s.t}
		&& T_{min} \leq T \leq T_{max} \\
		&&&  \delta T_{j,min} \leq \Delta T \leq \delta T_{j,max}  \\
		&&&  X_{t+1} = LSTM_s([X_{t-W}, X_{t-W+1} ... X_{t}])  \\
		\end{aligned}
\end{equation}
where $L_{sp}$ is the desired set-point, and $\Delta T_j$ corresponds to the maximum change in jacket temperature allowed between two sampling times. $LSTM_s$ is a surrogate model, developed to mimic the batch crystallization process. It has been employed to expedite the internal computations of a traditional MPC, facilitating practical online process control applications. In specific terms, $LSTM_s$ is trained using previously generated simulation data. A state tensor $X_t =[T_j(t), C_s(t), T(t), \bar{L}(t), \mu_0(t), \mu_1(t), \mu_2(t), \mu_3(t), M_T(t), t]$ for current and previous $W$ time-steps (i.e., $[X_{t-W}, ..., X_{t}]$) is considered as the input to this model. The output of the $LSTM_s$ model is the state predictions for the next time-step ($Y_{LSTM_s}$). Following the approach used for the LSTMc, $LSTM_s$ was constructed using 350K data points for the training set, 150K for the validation set, and an additional 100K for the testing set. The model validation results for three important states are presented in Figure~\ref{LSTM_validation}. These results demonstrate a strong alignment between the state evolution predictions and the actual values obtained from the PBM-based crystallization model. While we have chosen to exhibit the state evolution of just three states for simplicity, the NMSE for the full testing dataset is impressively low at $0.76 \times 10^{-3}$, indicating an exceptional overall model performance.

It is important to understand that while $LSTM_s$ and LSTMc are based on the LSTM architecture as shown in Figure~\ref{LSTMc_vs_LSTM}, they represent distinct models. Specifically, LSTMs takes in an input tensor comprising information about the state evolution for the current and previous $W$ time-steps and then predict the state values for the next time-step. On the contrary, LSTMc uses an augmented state tensor (which includes $L_{final}$ and $e$) as its model input to generate predictions for the jacket temperature and set-point error at the next time-step. In simpler terms, (a) $LSTM_s$ serves as a surrogate model for the crystallization process, generating state predictions that can be incorporated into an LSTM-MPC to calculate $T_j$ for the next time-step; and (b) LSTMc operates as a model-free, data-driven controller that processes state evolution and error dynamics from previous time-steps to directly output $T_j$ for the next time-step, thereby minimizing the set-point deviation at the end of the crystallization process. 

\begin{rmk}
The LSTMc and $LSTM_s$ models have the exact same network architecture. More specifically, both models are built using a 4-layer LSTM network, comprising 512 LSTM cells and resulting in approximately 125K parameters. Both models were trained on approximately 350K training points and required an average of 2 hours of training time on a single Nvidia A100 40GB GPU. To ensure a fair comparison, identical model architecture, quantities of training data, and training duration were considered. 
\end{rmk}

\subsection{Design of PI Controller}

To further compare against LSTMc, a PI controller was also designed. The control law for a PI controller can typically be defined as follows:

\begin{equation} \label{PI_controller}
	u(t) = K_c\left(e(t) + \frac{1}{\tau_I}\int_{0}^{t} e(t)\right)
\end{equation}
where $K_c$ is the controller gain, $\tau_I$ is the integral time constant, and $e(t)$ is the deviation between the current mean crystal size ($\bar{L}$) and set-point mean crystal size ($\bar{L_{sp}}$) at time $t$. Essentially, at each sampling step (i.e., every hour), the control law is used to determine the next input (i.e., jacket temperature, $T_j$). This predicted $T_j$ is applied to the crystallization system, which is represented by the PBM-based batch crystallizer model acting as a virtual experiment. The deviation $e(t)$ is then computed at that moment. This procedure is repeated until the end of the crystallization process (i.e., 24 hours), at which point the set-point tracking performance is evaluated. It is worth noting that ($K_c, \tau_I$) constitutes a specific set of tunable controller parameters that must be determined for each new set-point tracking case using methods such as trial-and-error, optimization techniques, or other tuning methods (e.g., Ziegler-Nichols and Cohen-Coon). In this work, several unique sets of ($K_c, \tau_I$) values were determined using a trial-and-error approach.

\section{Closed-Loop Simulations}
\subsection{Set-point Tracking Performance}

\begin{figure}[!ht]
	\begin{center}
		\centerline{\includegraphics[width=1\columnwidth]{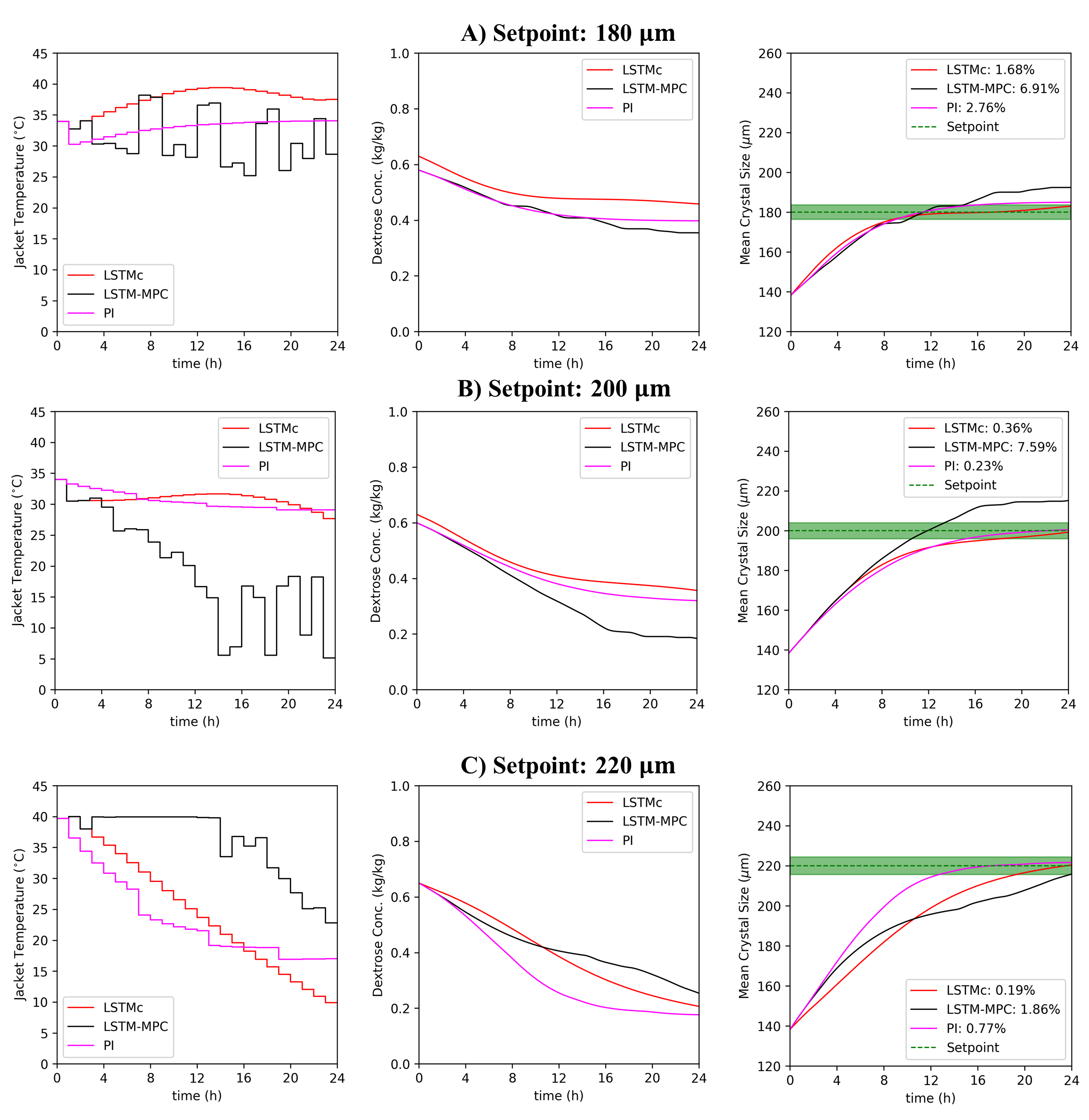}}
	\end{center}
	\caption{Closed-loop simulation results for the PI, LSTM-MPC, and LSTMc across various set-point cases.}
	\label{case_220}
\end{figure}

Figure~\ref{case_220} illustrates the performance of the LSTMc in three set-point tracking cases. The results clearly show that the LSTMc consistently delivers the smallest set-point deviation, which is always less than 2\%. Although the PI controller also shows a low set-point deviation, this was expected, as each PI controller set-point case was specifically tuned for that set-point. In other words, a tailored pair of ($K_c, \tau_I$) was calculated for each set-point case, as enumerated in Table~\ref{PI_tuning}. Conversely, LSTM-MPC shows a comparatively larger high set-point deviation in all three cases when compared to both the LSTMc and PI controller. 

\begin{table}[!ht]
	\centering
	\caption{The parameters for the PI controller across various set-point cases.}
	\label{PI_tuning}
	\begin{tabular}{lcc}
		\toprule
		set-point Cases & $K_c$ & $\tau_I$ \\
		\midrule
		180 $\mu$m & 0.01 & 1 \\
		200 $\mu$m & 0.005 & 0.1 \\
		220 $\mu$m & 0.01 & 0.05 \\
		\bottomrule
	\end{tabular}
\end{table}

Further analyzing the results reveal the following aspects; The LSTMc model is trained on more than 5000 different operating conditions (and tested on more than 2000 operating conditions), each with a unique $T_j$ profile, different initial dextrose concentration, and varying seeding conditions. Thus, LSTMc has the capability to interpolate seamlessly between different operating conditions, deriving a unique $T_j$ profile suitable for the given process conditions (e.g., set-point case A, B, or C). Also, by considering the state evolution of previous time-steps and error dynamics, LSTMc effectively learns the relationship between state evolution at a specific time $t$ and the corresponding error. Essentially, the training protocol based on the augmented state tensor enables LSTMc to integrate the knowledge of state evolution and error dynamics from a single operating condition with similar information from all other operating conditions, resulting in a comprehensive understanding of crystallization dynamics and their corresponding control actions. 

Regarding the developed LSTM-MPC framework, it involves several internal computations at each sampling time. Specifically, the LSTM-MPC uses $LSMT_s$ as an internal surrogate model to forecast the future state evolution over a control horizon of $H$. It then evaluates different $T_j$ values over this horizon to determine the trajectory of $T_j$ with the minimum set-point deviation. However, there is one drawback to the implementation of LSTM-MPC; The $LSMT_s$ model is trained to consider the state evolution over current and previous $W$ time-steps to predict states at the next time step. Although training and validation ensure that the NMSE value is less than $1\times 10^{-3}$ for future state predictions, the internal LSTM model is treated as a black-box optimization problem by the optimizer within the LSTM-MPC framework. Although this might not be an issue for small-sized ML models (i.e., with 2 to 3 process states), for complex chemical processes like in this case, the high-dimensional black-box nature of the $LSMT_s$ model may lead to LSTM-MPC traversing regions of non-convexity and encountering situations of multiple local minima points. As a result, the LSTM-MPC exhibits inferior set-point tracking performance compared to the LSTMc. 


\begin{rmk}
The LSTMc approach offers additional advantages, such as its computational efficiency. As it does not necessitate internal computations associated with exploring different scenarios with varying $T_j$ profiles, the average computational time for LSTMc is approximately 20 ms. This is 2000 times faster than that of the LSTM-MPC. Moreover, the computational bandwidth required for executing an offline trained LSTMc is substantially less than that needed for performing online computations using an LSTM-MPC. Considering the limited computational bandwidth available in current controller hardware across many chemical processes, deploying an ML-based MPC is not always feasible. However, a pretrained LSTMc model can be loaded onto a small micro-chip controller for implementation, similar to the case of implementing PI controllers. 
\end{rmk}

\begin{figure}[!ht]
	\begin{center}
		\centerline{\includegraphics[width=1\columnwidth]{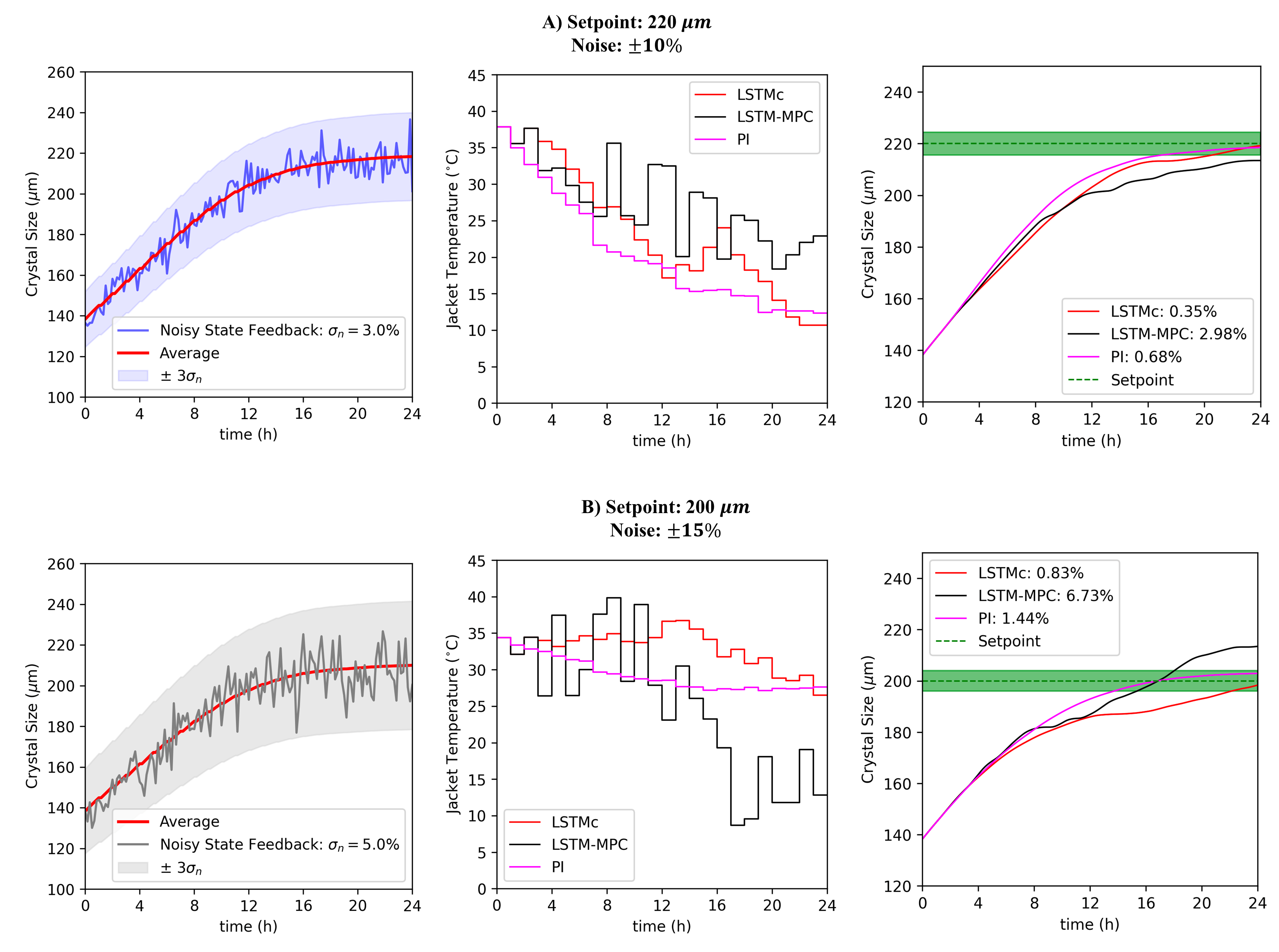}}
	\end{center}
	\caption{Closed-loop simulation results for PI, LSTM-MPC, and LSTMc for different set-point cases.}
	\label{noisy_cases}
\end{figure}

\subsection{Effect of Noisy Measurements}
Another frequent challenge in many industrial chemical processes is the presence of noisy measurements. These disrupt the feedback signal, potentially leading to suboptimal set-point tracking performance. To evaluate the resilience of LSTMc in the presence of measurement noise, artificial white Gaussian noise of varying magnitudes was introduced into the feedback signal of mean crystal size ($\bar{L}$) as follows: 

\begin{equation}
	\begin{gathered}
		p_i (x) = \frac{1}{\sigma_n\sqrt{2\pi}} 
		\exp\left( -\frac{1}{2}\left(\frac{x-\bar{p_i}}{\sigma_n}\right)^{\!2}\,\right) \\ 
		\bar{L}_{noise} = \bar{L} \pm p_i
	\end{gathered}
\end{equation}
where $p_i$ is the extent of noise in the signal, and $\sigma_n$ is the standard deviation of the Gaussian noise. Although the impact of noisy measurements for other state variables could also be considered, $\bar{L}$ is a significant tracking variable. As such, its noise can complicate the computation of set-point deviation, making it challenging to counteract.

Figure~\ref{noisy_cases} shows two scenarios, featuring $\pm 10\%$ and $\pm 15\%$ noise, where all three controllers are compared for their set-point tracking progress. Table~\ref{set-point_table} enumerates the controller performance for all the different cases, with and without process noise. Although the set-point tracking performance for all the controllers is slightly diminished compared to the case without noise, the LSTMc exhibits the least set-point deviation (under 2\%) even in the presence of considerable process noise. In the case of the PI controller, noisy measurements affect the error signal, which is directly linked to the control law, thereby influencing its performance. For the LSTM-MPC case, the addition of process noise in the state feedback would be incorporated into the input state tensor (i.e., $[X_{t-w}, ..., X_t]$) considered by the surrogate $LSTM_s$ model. This incorporation is likely to marginally impair its predictive abilities. These marginal errors may manifest as slightly less accurate state predictions, consequently influencing the internal computations of the LSTM-MPC as compared to situations devoid of any process noise (Figure~\ref{case_220}). 

The LSTMc showcases its unique capabilities due to the presence of numerous internal gates and weighting functions. These features allow the LSTMc to mitigate the effect of noise signals through its inherent pseudo-noise-filtering system. To illustrate, the computations as presented in Equation~\ref{LSTM_equations} equip the LSTMc to assimilate a source sequence, discard irrelevant information, assign greater importance to notable state changes, and transmit relevant information to the next step. When LSTMc processes noisy measurements, it is able to effectively filter the signal variations, primarily concentrating on the time-step coinciding with a significant process change, such as during a control action. Furthermore, even if certain variations from the noise signal are not eliminated by one of the gates, the sequential computations of the forget, input, and output gates succeed in progressively dampening the noise signal, thereby resulting in accurate set-point tracking performance. Additionally, due to these internal gates, the predicted $T_j$ profile from LSTMc exhibits greater dynamics in Figure~\ref{noisy_cases} compared to Figure~\ref{case_220}. This observation underlines LSTMc's proficiency in processing noisy signals and modifying the internal weighting scheme, resulting in a new and unique $T_j$ profile that allows for minimum set-point deviation. These capabilities, particularly notable in the LSTMc, are absent in traditional DNNs or RNNs. Therefore, LSTMc demonstrates a unique skill set in handling noisy data and ensuring accurate performance tracking.

\begin{table}[!ht]
	\centering
	\caption{Comparison of different set-point cases.}\label{set-point_table}
	\begin{tabular}{lcccc}
		\toprule
		set-point Cases & LSTMc (\%) & LSTM-MPC (\%) & PI (\%) \\
		\midrule
		180 $\mu$m & 1.68 & 6.91 & 2.76 \\
		180 $\mu$m (w/ Noise) & 1.78 & 7.01 & 2.81 \\
		200 $\mu$m & 0.36 & 6.73 & 0.23 \\
		200 $\mu$m (w/ Noise) & 0.83 & 7.59 & 1.44 \\
		220 $\mu$m & 0.19 & 1.86 & 0.68 \\
		220 $\mu$m (w/ Noise) & 0.35 & 2.98 & 0.77 \\
		\textit{Avg. Comp. Time} & \textit{20 ms} & \textit{40 s} & \textit{10 $\mu$s} \\
		\bottomrule
	\end{tabular}
\end{table}

\begin{rmk}
The dampening effect on process noise and variations, facilitated by the internal gates within LSTMc, can also be found in the recently developed time-series transformer (TST) models. These models utilize a combination of encoders and decoders, each incorporating a multiheaded attention mechanism. Essentially, this mechanism executes scaled-dot product calculations among various input tensors, allowing it to selectively focus on significant long-term (e.g., concentration evolution) and short-term (e.g., sudden change in temperature due to control actions) process changes. This selective focus is achieved by assigning higher attention scores to such occurrences. As such, the multiheaded attention mechanism serves as a dynamic filtering system. It adeptly manages process uncertainties and data noise by effectively diminishing weak correlations while emphasizing strong interactions between system states. Therefore, this novel TST-based hybrid model tackles not only operational process uncertainties but also those stemming from sensor measurements. As a result, it offers precise predictions of unknown time-varying parameters within a hybrid model. Recent studies have demonstrated the successful implementation of TST-based controllers \cite{sitapure2023exploring,sitapure2023crystalgpt}.  
\end{rmk}

\begin{figure}[!ht]
	\begin{center}
		\centerline{\includegraphics[width=0.85\columnwidth]{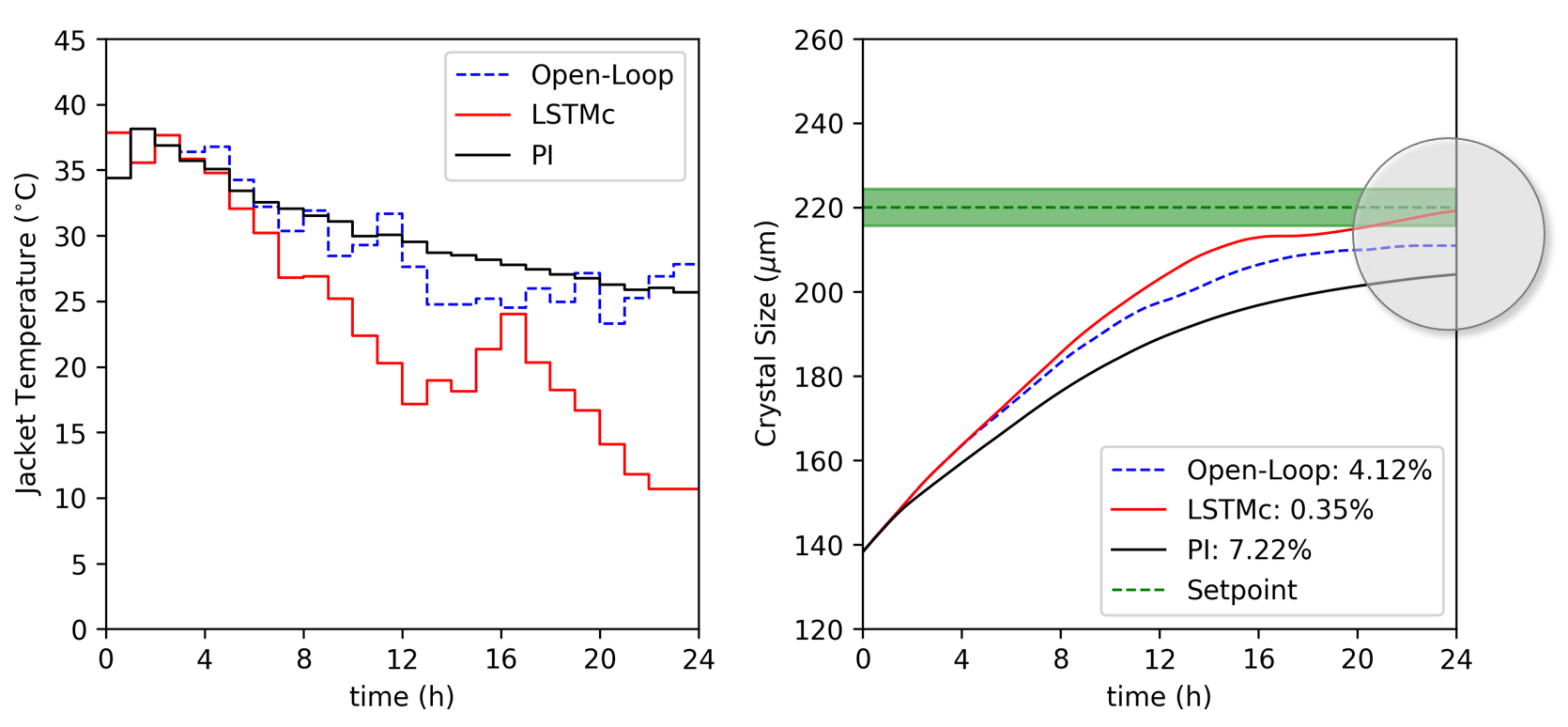}}
	\end{center}
	\caption{Closed-loop simulation results involving an inaccurately tuned PI controller and LSTMc under conditions of $\pm$10\% process noise.}
	\label{untuned_PI}
\end{figure}

\subsection{Seamless G2G Transferability}

Table~\ref{set-point_table} indicates that the performance of the PI controller holds up fairly well when compared to LSTMc. However, the PI controller we developed exhibits limited generalizability across different operating conditions, also known as poor G2G transferability. As illustrated in Table~\ref{PI_tuning}, it requires distinct tuning of process parameters for each unique operating case. To illustrate this issue, we refer to Figure~\ref{untuned_PI}, which depicts a set-point tracking case for a 220 $\mu$m setting. In this instance, we employed a PI controller that was initially tuned for a 180 $\mu$m setting. The figure clearly demonstrates that, although the PI controller attempts to steer the crystallization process in the right direction, it can only achieve approximately 200 $\mu$m, resulting in a set-point deviation of ~7\%. 

\begin{figure}[!ht]
	\begin{center}
		\centerline{\includegraphics[width=0.85\columnwidth]{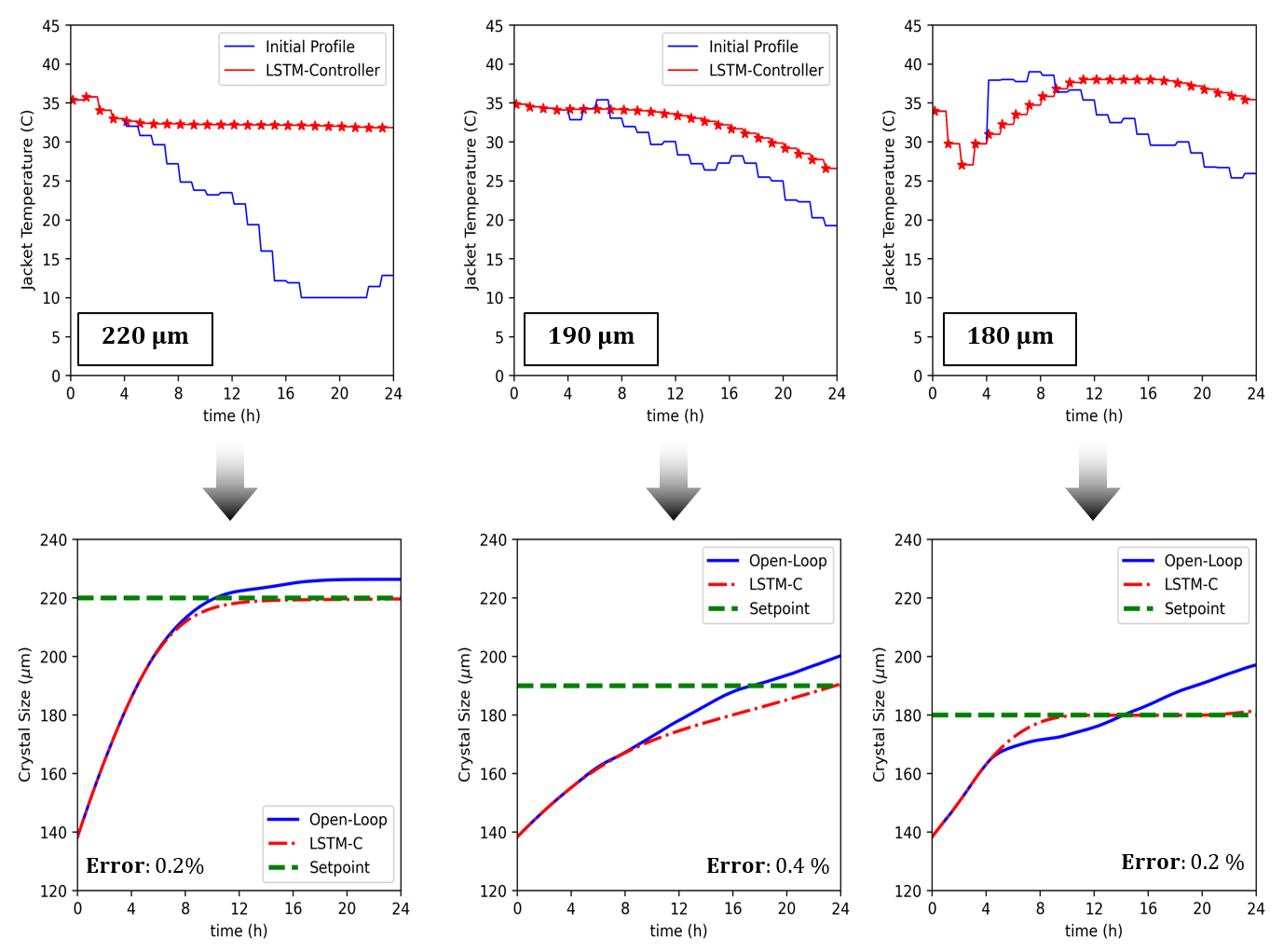}}
	\end{center}
	\caption{Closed-loop simulation results for different set-point cases, each with varying initial conditions.}
	\label{G2G_transfer}
\end{figure}

Contrary to the PI controller, Figures~\ref{untuned_PI} and \ref{G2G_transfer} demonstrate that the LSTMc exhibits seamless G2G transferability while maintaining a very low set-point deviation (less than 2\%). More interestingly, LSTMc can adeptly adapt to varying operating conditions and generate new $T_j$ profiles, ensuring efficient set-point tracking performance. For example, under the influence of process noise, LSTMc produces a dynamically changing cooling profile, as shown in Figure~\ref{untuned_PI}. In contrast, in the absence of process noise, LSTMc generates smooth cooling profiles, as depicted in Figure~\ref{G2G_transfer}. This uncanny cability of LSTMc can be explained as follows: LSTMc is trained on over 5000 operational conditions, encompassing state evolution and error dynamics. During the training phase, data from all these cases are fed into the training module, and LSTMc is tasked with determining the next input step ($T_j$) based on a particular state evolution over the previous $W$ time-steps and for a final crystal size (${L_{final}}$). However, because we combine data from all distinct operating conditions, the LSTMc learns to interpolate between these operating conditions. For instance, for operating condition 1, $T_j$ might be a step curve with decreasing values, while for condition 2, $T_j$ values might oscillate over time. In comparison, for operating condition 3, $T_j$ values may remain relatively constant, resulting in slow crystal growth and low nucleation rates. Essentially, as all permutations of operating conditions, state evolutions, and set-point cases are presented to the LSTMc during training, it naturally learns to interpolate among these conditions. As a result, it can produce a linearly independent weighted solution that is distinct from what a traditional PI or MPC can generate. 

Moreover, the LSTMc appears to integrate the simplicity of the PI controller with the G2G capabilities of an LSTM-MPC. This allows it to learn what can be described as a unified mapping function, effectively serving as a control law, as shown below: 

\begin{equation}
	\begin{gathered}
		u_{LSTM_c}(t) \approx \Lambda_1([X_{t-W}, X_{t-W+1}, ..., X_t]) + \Lambda_2\left(\int_{t_W}^{t} e\right)\\ 
		X_i = [T_j,C_s,T,\bar{L},\mu_0,\mu_1,\mu_2,\mu_3,M_T, time~|~{L_{final}}, e_i]
	\end{gathered} \label{LSTMc_control_law}
\end{equation}
where $\Lambda_1$ is a nonlinear mapping function that the LSTMc learns to correlate between the various system states and their evolution. Additionally, $\Lambda_2$ is another mapping that LSTMc learns, which takes into account the evolution of error dynamics over the current and previous $W$ time-steps. It is important to point out that Equation~\ref{LSTMc_control_law} does not present an explicit function learned by the LSTMc. Instead, it represents a hypothetical mapping function that the LSTMc might develop during its training as a process controller. More specifically, this equation form is inspired by the control law of a PI controller. Similar to the LSTMc, a PI controller also considers the the error dynamics from previous time-steps.

\begin{rmk}
	At first glance, the LSTMc might appear to resemble a reinforcement learning (RL)-based controller, which typically involves various components such as an actor, critic, and target network, often implemented as neural networks, like deep neural networks (DNNs) \cite{hubbs2020deep, yoo2021reinforcement, badgwell2018reinforcement}. In RL, the controller considers the system's current state at time $t$ and takes control actions through the actor-network to reach a desired target. The critic network then evaluates these actions and rewards, or penalizes them based on deviations from the set-point. While RL controllers have been explored in the literature, they present certain limitations. First, RL models only consider present state values and do not account for the evolution of previous $W$ states. This overlooks important information concerning state dynamics and error characteristics, which are critical for understanding process dynamics and assessing the time-varying impact of control actions on the system. Second, RL algorithms rely on separate actor, critic, and target networks, all implemented as individual DNNs trained simultaneously during episodic training. However, DNNs are known to perform suboptimally time-series modeling tasks when compared to LSTM networks. Additionally, the incorporation of DNNs in parallel and series can lead to complex, non-smooth control functions. Third, RL is primarily used in discrete systems such as AlphaGO, chess, and decision-induced video games \cite{lapan2018deep}, but its application in complex dynamic chemical processes necessities spatiotemporal discretization. This significantly increases the number of decision nodes in an RL algorithm, leading to a substantial computational burden \cite{doya2000reinforcement}. Therefore, model-order-reduction techniques are often employed before implementing RL, which can introduce potential plant-model mismatches due to the incomplete use of full-state information. Lastly, given the complexity of process systems with multiple state variables and extensive control spaces, episodic training in RL becomes computationally demanding compared to training an LSTM network. Both our experience with RL-based controllers and several literature studies suggest that episodic training for complex process systems with numerous state variables and large control spaces is considerably more computationally costly than training an LSTM network \cite{bangi2021deep, liu2019sequence}. 
\end{rmk}

\section{Conclusions}

Despite the commendable performance of existing controllers in regulating complex chemical processes, they possess certain limitations. These include: (a) Traditional PI controllers show poor G2G transferability, often requiring bespoke tuning for different set-point tracking cases; (b) The application of an MPC framework involves multiple resource-intensive steps such as training and testing of a surrogate model, formulating an internal optimization problem, and tuning the MPC; and (c) Frequently, the use of black-box-based ML models in MPC can lead to issues like traversing through areas of infeasibility, non-convexity, and local minima. Moreover, the existing industrially available controller hardware lacks the computational bandwidth needed for rapid online computations required by an MPC. To address these challenges, we developed a first-of-a-kind LSTM controller (LSTMc) - a model-free, data-driven control framework. The LSTMc employs an augmented input tensor, which includes information on state evolution and error dynamics over current and previous $W$ time-steps, to predict the manipulated input at the next step ($u_{t+1}$). We demonstrated this proposed framework using a case study of batch crystallization of dextrose, where the jacket temperature ($T_j$) was the manipulated input and the mean crystal size ($\bar{L}$) was the desired output requiring set-point tracking. A PI controller and an LSTM-MPC were designed for comparative purposes across several different set-point tracking cases. In all cases, the LSTMc consistently exhibited the least set-point deviation (less than 2\%), which is three times lower than that of LSTM-MPC. Interestingly, the LSTMc maintained this superior performance across all different set-points, even when 10\% to 15\% noise was added to sensor measurements, demonstrating seamless G2G transferability. We attribute the remarkable performance of the LSTMc to three primary factors: (a) The LSTMc learns not only the relationship between system states, but also the correlation between states and error dynamics; (b) The presence of various internal gates that dynamically weigh different input sequences based on their relevance, enabling the LSTMc to focus on time-steps with significant changes (e.g., a control action); and (c) These internal gates dampen process noise and act as pseudo noise filters. In a nutshell, LSTMc presents a highly promising alternative for controller design. It adeptly leverages the availability of process data and the efficient use of sequential ML models, resulting in superior controller performance with straightforward implementation.

\section{Acknowledgments}
Financial support from the Artie McFerrin Department of Chemical Engineering, and the Texas A\&M Energy Institute is gratefully acknowledged.

\newpage

\bibliographystyle{unsrt}  
\bibliography{LSTMc_bib}  
\end{document}